\documentclass[10pt]{article}
\usepackage{graphicx}
\usepackage{latexsym,graphicx}
\usepackage{float}
\usepackage{amsmath}
\usepackage{amsfonts}
\usepackage{amssymb}
\usepackage{epstopdf}

 \newcommand{\be}{\begin{equation}}
 \newcommand{\ee}{\end{equation}}
 \newcommand{\bea}{\begin{eqnarray}}
 \newcommand{\eea}{\end{eqnarray}}
\DeclareGraphicsExtensions{.eps}
\usepackage[left=2.5cm,top=2.5cm,right=2.5cm,bottom=2.5cm]{geometry}

\begin{document}
	\begin{center}
		\large{\bf{A flat FRW model with dynamical $\Lambda$ as function of matter and geometry}} \\
		\vspace{5mm}
		\normalsize{ Anirudh Pradhan$^1$, De Avik $^2$, Tee How Loo$^3$, Dinesh Chandra Maurya$^4$}\\
		\vspace{5mm}
		\normalsize{$^{1}$Department of Mathematics, Institute of Applied Sciences \& Humanities, GLA University,\\ 
                 Mathura -281 406, Uttar Pradesh, India} \\
		\vspace{5mm}
		\normalsize{$^{2}$Department of Mathematical and Actuarial Sciences, Universiti Tunku Abdul Rahman, Jalan Sungai Long, 
		43000 heras, Malaysia} \\
		\vspace{5mm}
		\normalsize{$^{3}$Institute of Mathematical Sciences, University of Malaya, 50603 Kuala Lumpur, Malaysia  } \\
		 \vspace{5mm}
                 \normalsize{$^{4}$Department of Mathematics, Faculty of Engineering \& Technology, IASE (Deemed to be University), 
	Sardarshahar-331 403, Rajsthan, India} \\ 
	 \vspace{2mm}
        $^1${Email:pradhan.anirudh@gmail.com}\\ 
	 \vspace{2mm}
	$^2${Email:de.math@gmail.com}\\
	 \vspace{2mm}
	$^3${Email:looth@um.edu.my}\\
	\vspace{2mm}
       $^4${Email:dcmaurya563@gmail.com}\\
	\end{center}
	\vspace{10mm}

\begin{abstract}
 We revisit the evolution of the scale factor in a flat FRW spacetime with a new generalized decay rule for the dynamic $\Lambda$-term under 
modified theories of gravity. It analyses certain cosmological parameters and examines their behaviours in this generalized setting which 
includes several decay laws in the literature. We have also obtained observational constraints on various model parameters and estimated 
the present values of cosmological parameters $\{\Omega_{m_0}$, $\Omega_{\Lambda_0}$, $q_{0}, t_{0}$, $\omega_{0}\}$ and have discussed 
with various observational results. Finite time past and future singularities in this model are also discussed.  
\end{abstract}

 \smallskip
%\date{}
%\maketitle
\smallskip
%%%%%%%%%%%%%%%%%%%%%%%%%%%%%%%%%%%%%%%%%%%%%%%%%%%%%%% SECTION 1 %%%%%%%%%%%%%%%%%%%%%%%%%%%%%%%%%%%%%%%%%%%%%%%%%%%%%%%%%%%%%
\section{\textbf{Introduction}}
The idea of the cosmological constant $\Lambda$ was first introduced by Einstein in 1917 to obtain a static universe because of 
Einstein's field equations in general relativity were showing a non-static universe i.e. the universe has either contracting or expanding 
nature and at that time observational studies supported an static universe due to presence limited observational data sets. Therefore, 
he revised his original field equations proposed in 1915 to remove this feature of size-changing. But the discovery of Hubble in 1929 
suggested that our universe is in expanding nature which was already obtained by the Friedmann-Lamatire-Rabortson-Walker in 1922 in 
their solution of Einstein's field equations using the same metric and this finding was once removed the newly added cosmological 
term $\Lambda$.\\

After the discovery of accelerating universe in 1998, the cosmological term came back in its place but at this time it did not prevent 
the expanding nature of universe, but in order to explain why it did not continues expanding nature with a constant rate. Although it 
should be possible for the known amount of matter present in the universe to exert adequate gravitational force to slow down the expansion, 
initiated with Big Bang. However, modern studies of standard candles showed that instead of decreasing, the rate at which it is expanding 
actually increases. The new generation of cosmologists proposes the existence of an unknown, exotic, homogeneous energy density
with high negative pressure referred to as dark energy (DE) to explain these observational findings. And the idea of cosmological term 
$\Lambda$ is the simplest contributing factor of this dark energy.\\

We still face the critical question of whether $\Lambda$ is always a fundamental constant or a mild dynamical variable constant after 
almost two decades. It turns out that the selection of $\Lambda$ as dark energy faces some serious issues in terms of (a) 
the ``old" cosmological constant problem (or fine-tuning problem) and (b) the cosmic coincidence problem, if we consider it as a constant 
rather than a specific dynamical variable. Moreover, the overall fit to the cosmological observables SNIa+BAO+H(z)+LSS+BBN+CMB has also been 
shown to support the class of unique dynamic models of the so-called fundamental constant $\Lambda$ as well.  
Dynamic model of $\Lambda$ may ultimately be necessary not only as a new epitome to improve the EFE, but also phenomenologically 
to unwind a number of tensions between the concordance model and the observation that is indicating towards DE. There are innumerable 
$\Lambda(t)$ laws available in the literature although all of which might not be viable under theoretical and observational ground and 
they might not follow covariant action, but they are still interesting to study from phenomenological ground. For example, 
Carvalho {\it et al.} \cite{ref1} and Waga \cite{ref2} first considered the case $\Lambda\propto (\frac{\dot{a}}{a})^2$ as a generalization 
of Chen and Wu's result advocating the possibility that the (effective) cosmological constant $\Lambda$ varies in time as $a^{-2}$. 
The case $\Lambda\propto \rho$ was first studied by Viswakarma \cite{ref3} and the case $\Lambda\propto \frac{\ddot{a}}{a}$ 
by Arbab \cite{ref4}. Ray {\it et al.} \cite{ref5} and Viswakarma \cite{ref6} showed that if we consider these three cases separately, 
the free parameters satisfy certain relations. Recently, Pan \cite{ref7} investigated dynamical $\Lambda$ as a function of several form of 
polynomials of $H$ and $\dot{H}$, discussing the finite time future singularities for each case. \\ 

Although several studies being done on these particular phenomenological models in the past, with more powerful observational analysis 
of the present days, most of these are detected to be suffering from certain theoretical and observational limitations. The 
$\Lambda\propto H^2$ and $\Lambda\propto \dot{H}$ cases are largely overruled now since the corresponding linear growth of cosmic 
perturbations is strongly disfavored by the observational data \cite{ref8}, \cite{ref9}. If $\Lambda \propto \rho$ alone, one can 
trivially absorb it to the constant and hence there will be no dark energy scenario in the cosmological evolution. This is indeed very 
clear for a perfect fluid with equation of state $p=\omega \rho$. The introduction of $\Lambda\propto \rho$ effectively modifies the 
proportionality constant $\omega$ and the model becomes equivalent to one with merely a new fluid only. We tried to overcome this 
issue by introducing a new decay law which combines both matter and geometry into the dynamical cosmological term. Moreover, we can 
show that a motivation towards the present decay law of $\Lambda$ can be drawn from the f(R)-gravity or equivalently from Brans-Dicke 
scalar-tensor theory  derived from a covariant action.   \\

In cosmology, the fundamental variables are either constructed from a spacetime metric directly (geometrical), or depend upon properties 
of physical fields. Naturally the physical variables are model-dependent, whereas the geometrical variables are more universal. The most 
popular geometric variables are the Hubble constant $H$ and the deceleration parameter $q$ whose values depend on the first and second 
time derivatives of the scale factor, respectively. We use these two parameters extensively to resolve the mystery of the inflationary 
universe. However, a more sensitive discrimination of the expansion rate and hence dark energy can be performed by considering a third 
time derivative in the general form for the scale factor of the Universe 
$$ a(t)=a(t_0)+\dot{a}|_0(t-t_0)+\frac{\ddot{a}|_0}{2}(t-t_0)^2+\frac{\dddot{a}|_0}{3}(t-t_0)^2+\dots  $$
This led Sahni {\it et al.} \cite{ref10} and Alam {\it et al.} \cite{ref11} to introduce the statefinder parameters, a dimensionless 
pair $\{r,s\}$ from the scale factor and its third order time derivatives in concordance cosmology. In \cite{ref12} this result is 
generalized to show that the hierarchy, $A_n=\frac{a^{(n)}}{aH^n}$, can take into account the deceleration parameter, the statefinder 
parameters, the snap parameter, the lerk parameter etc and further it can be expressed in terms of the deceleration parameter or the 
matter density parameter.\\ 

In observational cosmology, the beginning of the 21st century saw yet another significant milestone. BOOMERanG in the year 2000 collaboration 
(Balloon Measurements of Millimetric Extragalactic Radiation and Geophysics), a research on cosmic microwave history (CMB) a cosmos of flat 
spatial geometry (\cite{ref13}, \cite{ref14}) was recorded using balloon-borne instruments. 
The MAXIMA (Millimeter Anisotropy Experiment Imaging Array) collaboration \cite{ref15,ref16}, DASI 
(Degree Angular Scale Interferometer) \cite{ref17}, CBI (Cosmic Background Imager) \cite{ref18} and WMAP (Wilkinson Microwave Anisotropy Probe) 
\cite{ref19} have also reported a similar result.  These observations were an significant landmark in modern cosmology; assuming astronomy's 
value of $\Omega M\sim0.3, $ the data directly pointed to a positive cosmological constant with a contribution of energy density of the order 
of $\Omega \Lambda\sim0.7.$ The observations from the supernova probes and the Hubble Space Telescope were ideally matched. 
The precise values of these parameters $\Omega_M$ and $\Omega_\Lambda$ as obtained by Sievers {\it et al.} \cite{ref18} and Spergel 
{\it et al.} \cite{ref20} are $[0.34\pm0.12,0.67^{+0.10}_{-0.13}]$ and $[0.249^{+0.024}_{-0.031},0.719^{+0.021}_{-0.029}]$, respectively. 
More recently, the Lyman-$\alpha$ forest measurement of the baryon acoustic oscillations (BAO) by the Baryon Oscillation Spectroscopic Survey 
preferred an even smaller value of the matter density $\Omega_M$ than that obtained by cosmic microwave background data \cite{ref21}. We show 
that our model perfectly adapts to these observational data.\\

The present paper is organized in the following format: a brief review of literature is given in section-1, section-2 contains field 
equations with cosmological term $\Lambda$ and some specific cosmological solutions like scale factor $a(t)$, Hubble function $H(t)$, 
energy density parameters $\Omega_{m}$ and $\Omega_{\Lambda}$. An observational constraints on various model parameters using available 
observational data sets like $H(z)$, union 2.1 compilation of SNe Ia data sets and Joint Light Curve Analysis (JLA) etc. by applying 
$R^{2}$-test formula are obtained and discussed with various observational results in section-3. Some finite time singularities of the 
current model are discussed in the section-4 and in section-5 we have discussed some more cosmological parameters like statefinder with 
deceleration parameters. Finally conclusions are given in section-6.

%%%%%%%%%%%%%%%%%%%%%%%%%%%%%%%%%%%%%%%%%%%%%%%%%%%SECTION 2 %%%%%%%%%%%%%%%%%%%%%%%%%%%%%%%%%%%%%%%%%%%%%%%%%%%%%%%%%%%%%%%%%%
\section{\textbf{Field equations with $\Lambda(t)$}}
We consider an action \cite{ref22}

\[S=\frac{1}{16\pi G}\int f(R) \sqrt{-g}d^4x +\int L_m\sqrt{-g}d^4x,\]

where $f(R)$ is an arbitrary function of the Ricci scalar $R$, $L_m$ is the matter Lagrangian density, and we define the stress-energy 
tensor of matter as $T_{\mu\nu}=-\frac{2}{\sqrt{-g}}\frac{\delta(\sqrt{-g}L_m)}{\delta g^{\mu\nu}}$. \\

Assuming that the Lagrangian density of matter $L m$ depends only on the components of the metric tensor $g {\mu\nu}$ and not on its components,
we obtain derivatives, 
$$ T_{\mu\nu}=g_{\mu\nu}L_m-2\frac{\partial L_m}{\partial g^{\mu\nu}}.$$

By varying the action $S$ of the gravitational field with respect to the metric tensor components $g^{\mu\nu}$ and using the least 
action principle we obtain the field equation

\be f_R(R)R_{\mu\nu}-\frac{1}{2}f(R)g_{\mu\nu}+(g_{\mu\nu}\Box-\nabla_\mu\nabla_\nu)f_R(R)=8\pi G \, T_{\mu\nu},\label{frt}	\ee

where $\Box=\nabla^2=\nabla_\mu\nabla^\mu$ represents the d'Alembertian operator and $f_R=\frac{\partial f(R)}{\partial R}$. EFE can 
be reawakened by putting $f(R)=R$. The contracted equation is given by

\be Rf_R(R)-2f(R)+3\Box f_R(R)=8\pi GT,\ee
which transforms (\ref{frt}) to
\be R_{\mu\nu}-\frac{R}{2}g_{\mu\nu}+\frac{1}{6f_R(R)}[2kT+f(R)+Rf_R(R)]g_{\mu\nu}=8\pi G\, T^{\text{eff}}_{\mu\nu},\label{frft}\ee

$T^{\text{eff}}_{\mu\nu}=\frac{1}{f_R(R)}[T_{\mu\nu}+\frac{1}{8\pi \, G}\nabla_\mu\nabla_\nu f_R(R)]$.\\

 The Friedmann-Robertson-Walker (FRW) metric is given by the line element
 
\be ds^2=-dt^2+a^2(t)\left[\frac{dr^2}{1-kr^2}+r^2(d\theta^2+\sin^2\theta d\phi^2)\right],\label{frw}\ee

where the curvature parameter $k=-1,0,1$ for open, flat and closed models of the universe, respectively and $a(t)$ is the scale factor 
shaping the universe. Since current observations from CMB detectors such as BOOMERanG, MAXIMA, DASI, CBI and WMAP confirm a spatially 
flat universe, $k=0$. Hence the Ricci scalar becomes $R=6((\frac{\dot{a}}{a})^2+\frac{\ddot{a}}{a})$. Comparing the equation (\ref{frft}) 
with the Einstein's field equations with cosmological constant

\be R_{\mu\nu}-\frac{R}{2}g_{\mu\nu}+\Lambda g_{\mu\nu}=8\pi G T_{\mu\nu},\label{efe}\ee 

we conclude that the dynamic cosmological term $\Lambda(t)$ is possibly a function of $\frac{\ddot{a}}{a}, \,(\frac{\dot{a}}{a})^2$ and 
$\rho$ and we consider the simplest model as a linear combination 

\be \Lambda=l\frac{\ddot{a}}{a}+\lambda \left(\frac{\dot{a}}{a}\right)^2+4\pi G\eta\rho,\label{ourlambda}\ee
where $l$, $\lambda$ and $\delta$ are arbitrary constants.\\

As usual, the Friedmann and Raychaudhuri equations for flat FRW line element are given by 

\be 3\frac{\ddot{a}}{a}+4\pi G(1+3w)\rho=\Lambda, \label{fried}\ee

\be \frac{\ddot{a}}{a}+2\left(\frac{\dot{a}}{a}\right)^2-4\pi G(1-w)\rho=\Lambda, \label{ray}\ee

where $w = \frac{p}{\rho}$. \\

From Eqs. (\ref{fried}) and (\ref{ray}) we obtain

\be\left(\frac{\dot{a}}{a}\right)^2-\frac{8\pi G\rho}{3}=\frac{\Lambda}{3}.\label{3}\ee

Using Eq. (\ref{ourlambda}), in terms of Hubble parameter $H=\frac{\dot{a}}{a}$, Eq. (\ref{3}) transforms into 

\be 3H^2-8\pi G\rho= l \dot{H}+(l+\lambda)H^2+4\pi G \eta \rho.\ee

Therefore, 
\be4\pi G\rho=-\frac{l}{2+\eta}\dot{H}+\frac{3-l-\lambda}{2+\eta}H^2\label{rho}\ee 
and
\be\Lambda=\frac{2l}{2+\eta}\dot{H}+\frac{2l+2\lambda+3\eta}{2+\eta}H^2\label{lambda}.\ee

From Eq. (\ref{ray}) we obtain 

\be \frac{\dot{H}}{H^2}=-\frac{(1+w)(3-l-\lambda)}{(2+\eta)-l(1+w)},\label{hdot}\ee

which gives us 

\be H = \frac{H_0[(2+\eta)-l(1+w)]}{H_0(1+w)(3-l-\lambda)(t-t_0)+(2+\eta)-l(1+w)}.\ee

Integrating again, we obtain

\be \frac{a(t)}{a_0}=\left[1+\frac{H_0(1+w)(3-l-\lambda)}{(2+\eta)-l(1+w)}(t-t_0)\right]^{\frac{(2+\eta)-
l(1+w)}{(1+w)(3-l-\lambda)}}.\label{scalefac}\ee 

Therefore, the vacuum energy density parameter ($\Omega_\Lambda=\frac{\Lambda}{3H^2}$) and the cosmic matter density parameter 
($\Omega_m=\frac{8\pi G\rho}{3H^2}$) are calculated as

\be 3\Omega_\Lambda=\frac{2l+2\lambda+3\eta-3l(1+w)}{(2+\eta)-l(1+w)}\label{omegalambda}\ee
and
\be 3\Omega_m=2\frac{3-l-\lambda}{(2+\eta)-l(1+w)}\label{omegam},\ee
respectively, which confirms $$\Omega_m+\Omega_\Lambda=1.$$

%%%%%%%%%%%%%%%%%%%%%%%%%%%%%%%%%%%%%%%%%%%%%%% SECTION 3 %%%%%%%%%%%%%%%%%%%%%%%%%%%%%%%%%%%%%%%%%%%%%%%%%%%%%%%%%%%%%%%%%%%%%%%%%%%
\section{\textbf{Observational constraints}}
The Hubble parameter $H$ is one of the key parameter in observational cosmology as well as in theoretical cosmology and it measures 
expansion rate of the universe. In our paper, we have obtained the Hubble function using Eqs. $(14)$ \& $(15)$ as

\begin{equation}\label{18}
H(t)=H_{0}\left(\frac{a_{0}}{a(t)}\right)^{\frac{(1+\omega)(3-l-\lambda)}{2+\eta-l(1+\omega)}} 
\end{equation}

Several observational data sets are available in terms of redshift $z$ and hence, to obtain constraints on various model parameters 
we have to convert the above relationship in terms of the redshift. Therefore, using the relation $\frac{a_{0}}{a(t)}=1+z$ between 
scale factor $a(t)$ and redshift $z$, we obtain:

\begin{equation}\label{19}
H(z)=H_{0}(1+z)^{\frac{(1+\omega)(3-l-\lambda)}{2+\eta-l(1+\omega)}} 
\end{equation}

We can also obtain the expression for luminosity distance $D_{L}$ and apparent magnitude $m(z)$ in the purpose of observational 
constraints (Union 2.1 compilation of SNe Ia data sets and Joint light curve analysis data sets) respectively as

\begin{equation}\label{20}
D_{L}=\frac{c}{H_{0}}\frac{2+\eta-l(1+\omega)}{2+\eta+(\lambda-3)(1+\omega)}(1+z)^{\frac{4+2\eta+(\lambda-l-3)(1+\omega)}
{2+\eta-l(1+\omega)}}
\end{equation}
and
\begin{equation}\label{21}
m(z)=16.08+5\times log10\left(\frac{1}{0.026}\frac{2+\eta-l(1+\omega)}{2+\eta+(\lambda-3)(1+\omega)}(1+z)^{\frac{4+2\eta+(\lambda-l-3)
(1+\omega)}{2+\eta-l(1+\omega)}} \right) 
\end{equation}

Now, we consider the $29$ Hubble data sets $H(z)$ \cite{ref23}-\cite{ref29}, $580$ data sets of apparent magnitude $m(z)$ from 
union 2.1 compilation of SNe Ia data sets \cite{ref30} and $51$ data sets of $m(z)$ from Joint Light Curve Analysis (JLA) data sets 
\cite{ref31} and using the $R^{2}$-test formula, we obtain the best fit values of various model parameters $l$, $\lambda$, $\eta$, 
$\omega$ and $H_{0}$ for the best fit curve of Hubble function $H(z)$ and apparent magnitude $m(z)$. The case $R^{2}=1$ means the curve 
of the Hubble function is exact match with observational curve and using this technique of fitting we have obtained the constraints on 
model parameters for maximum $R^{2}$ values. The best fit values of $l$, $\lambda$, $\eta$, $\omega$ and $H_{0}$ for the various 
data sets are mentioned in below Table-1 \& 2 with their maximum $R^{2}$ value and root mean square error (RMSE).\\

%%%%%%%%%%%%%%%%%%%%%%%%%%%%%%%%%%%%%%%%%%%%%%%%%%%%%
\begin{table}[ht]
	\centering
	{\begin{tabular}{ccccc} 
	%{ccccc@rrrrrrrrrrrrrrrrrrrrrrrrrrrrrr}
			\hline\hline
			Parameters(For $\omega=0$) & $H(z)$ & SNe Ia & JLA \\
			\hline
			$l$ & $1.435$ & $0.1707$ & $0.2171$\\
			
			$\lambda$ & $0.8113$ & $0.2855$ & $0.7471$\\
			
			$\eta$ & $0.1044$ & $1.119$ & $0.8394$\\
			
			$H_{0}$ & $59.22$ & $-$ & $-$\\
			
			$R^{2}$ & $0.9115$ & $0.9938$ & $0.9214$\\
			
			$RMSE$ & $13.2672$ & $0.2486$ & $0.2782$\\
			
			\hline\hline
	\end{tabular}}
	\caption{Case-I dusty universe: The best fit values for the model parameters $l$, $\lambda$, $\eta$ and $H {0}$ are used in
                the best fit curve of the Hubble function $H(z)$ and the apparent magnitude $m(z)$ with different observational 
                 data sets ($H(z)$, JLA, SNe Ia) with confidence level of constraints of $95 \%$. }
\end{table}
%%%%%%%%%%%%%%%%%%%%%%%%%%%%%%%%%%%%%%%
\begin{table}[ht]
	\centering
	{\begin{tabular}{ccccc@rrrrrrrrrrrrrrrrrrrrrrrrrrrrrr}
			\hline\hline
			Parameters & $H(z)$ & SNe Ia & JLA \\
			\hline
			$\omega$ & $0.1256$ & $0.1156$ & $0.1218$\\
			
			$l$ & $1.764$ & $0.1881$ & $0.1137$\\
			
			$\lambda$ & $0.7351$ & $0.8632$ & $0.9196$\\
			
			$\eta$ & $0.4916$ & $0.7292$ & $0.9696$\\
			
			$H_{0}$ & $59.83$ & $-$ & $-$\\
			
			$R^{2}$ & $0.9114$ & $0.9938$ & $0.9214$\\
			
			$RMSE$ & $13.5632$ & $0.2488$ & $0.2812$\\
			
			\hline\hline
	\end{tabular}}
	\caption{Case-II $0<\omega<\frac{1}{3}$: The best fit values for the model parameters $l$, $\lambda$, $\eta$ and $H {0}$ are used for
                  the best fit curve for the Hubble function $H(z)$ and the apparent magnitude $m(z)$ for different observational data 
                   sets ($H(z)$, JLA, SNe Ia) for confidence level of boundaries of $95\%$. }
\end{table}
%%%%%%%%%%%%%%%%%%%%%%%%%%%%%%%%%%%%%%%
\begin{table}[ht]
	\centering
	{\begin{tabular}{ccccc}
	%{ccccc@rrrrrrrrrrrrrrrrrrrrrrrrrrrrrr}
			\hline\hline
			Parameters (for $\omega=0$) & $H(z)$ & SNe Ia & JLA \\
			\hline
			$\Omega_{m}$ & $0.7506224480$ & $0.5752015285$ & $0.5175609198$\\
			
			$\Omega_{\Lambda}$ & $0.2493775520$ & $0.4247984713$ & $0.4824390803$\\
			
			$q$ & $0.125933672$ & $-0.1371977072$ & $-0.2236586203$\\
			
			$t_{0}$ & $14.56595745$ & $18.74417427$ & $20.62684904$\\
			
			$r$ & $0.1576522515$ & $-0.09955128548$ & $-0.1236122635$\\
			
			$s$ & $0.7506224480$ & $0.5752015286$ & $0.5175609198$\\
			
			\hline\hline
	\end{tabular}}
	\caption{Case-I dusty universe: The present values of various parameters obtained for different data sets.}
\end{table}
%%%%%%%%%%%%%%%%%%%%%%%%%%%%%%%%%%%%%%%
\begin{table}[ht ]
	\centering
	{\begin{tabular}{ccccc}
	%{ccccc@rrrrrrrrrrrrrrrrrrrrrrrrrrrrrr}
			\hline\hline
			Parameters & $H(z)$ & SNe Ia & JLA \\
			\hline
			$\omega$ & $0.1256$ & $0.1156$ & $0.1218$\\
			
			$\Omega_{m}$ & $0.6598930470$ & $0.5156609542$ & $0.4613334442$\\
			
			$\Omega_{\Lambda}$ & $0.3401069530$ & $0.4843390457$ & $0.5386665557$\\
			
			$q$ & $-0.3115722925$ & $-0.2331273858$ & $-0.3113554396$\\
			
			$t_{0}$ & $14.54821996$ & $18.53672071$ & $20.41654688$\\
			
			$r$ & $-0.1174177056$ & $-0.1244306298$ & $-0.1174710201$\\
			
			$s$ & $0.4589518050$ & $0.5112484095$ & $0.4590963736$\\
			
			\hline\hline
	\end{tabular}}
	\caption{Case-II of $0<\omega<\frac{1}{3}$: The present values of various parameters obtained for different data sets.}
\end{table}
%%%%%%%%%%%%%%%%%%%%%%%%%%%%%%%%%%%%%%%%%%%%%%%%%%%%%%%%%%%%%%%%%%%%%%%%%%%%%%%%
Using the above estimated values of various parameters (as mentioned in Table-1 \& 2) we have calculated the values of matter 
energy density parameter $\Omega_m$, dark energy density parameter $\Omega_\Lambda$, deceleration parameter $q$, age of the present 
universe and statefinder parameters $r$, $s$ for two types of universes one is dusty universe and second is the case of $0<\omega<\frac{1}{3}$ 
which are mentioned in Table-3 \& 4 given below.\\

Table-3 \& 4 shows the present values of various cosmological parameters $\{\Omega_{m_0},\Omega_{\Lambda_0}, q_{0}, t_{0},r_{0},s_{0}\}$ 
and we see that for the deceleration in expansion of the universe dark energy parameter must be $\Omega_{\Lambda_0}>0.2493775520$ and matter 
energy density parameter $\Omega_{m_0}<0.7506224480$. In our model we have calculated the present value of deceleration parameter $q_{0}$ 
lies between $-0.3115722925\leq q_{0}\leq0.125933672$. From these results we concludes that our present universe is in accelerating phase.\\

%%%%%%%%%%%%%%%%%%%%%%%%%Figure 1 %%%%%%%%%%%%%%%%%%%%%%%%%%%%%%%%%%%%%%%%%%%%
\begin{figure}
	\centering
	a.\includegraphics[width=6cm,height=6cm,angle=0]{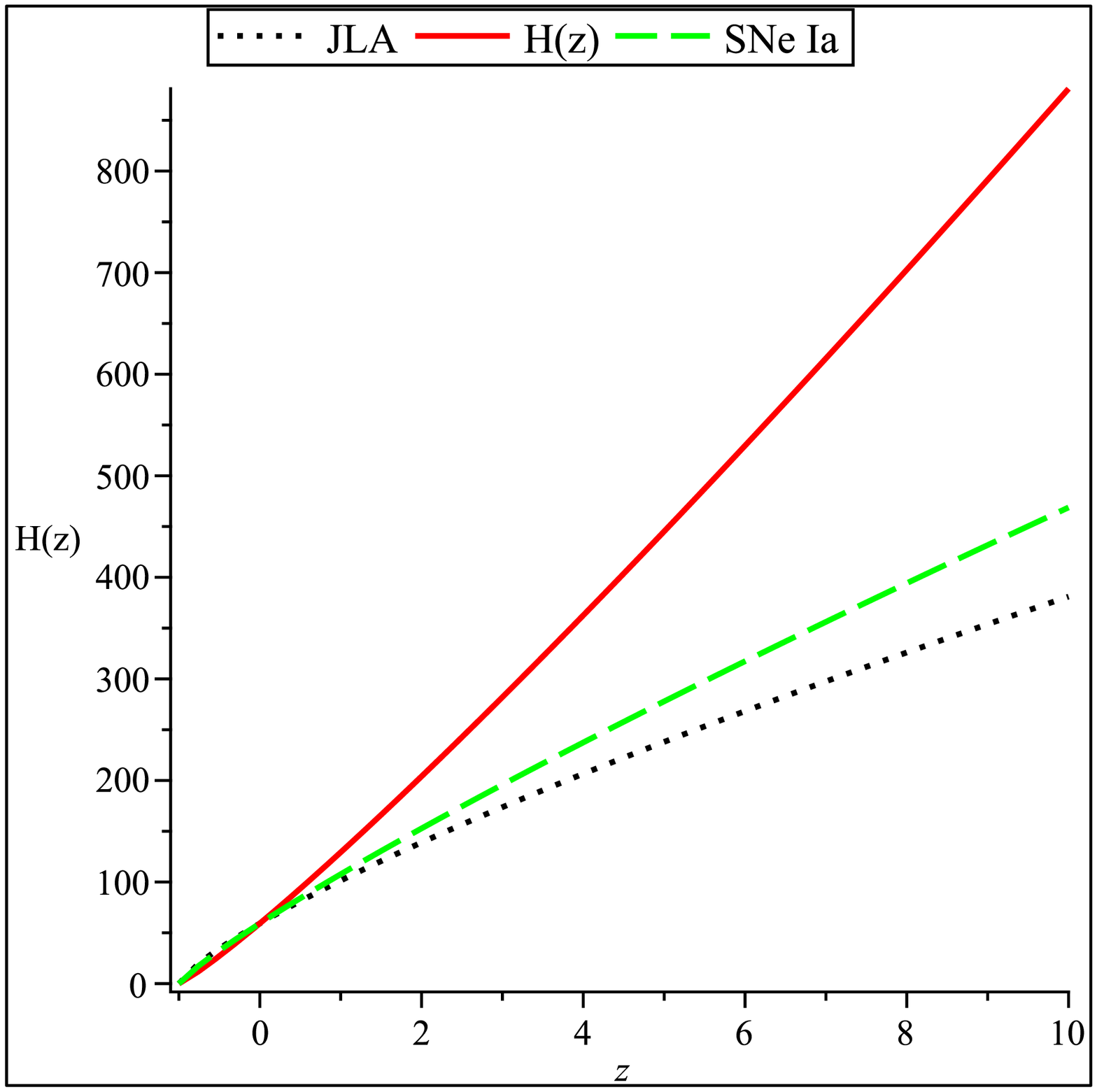}
	b.\includegraphics[width=6cm,height=6cm,angle=0]{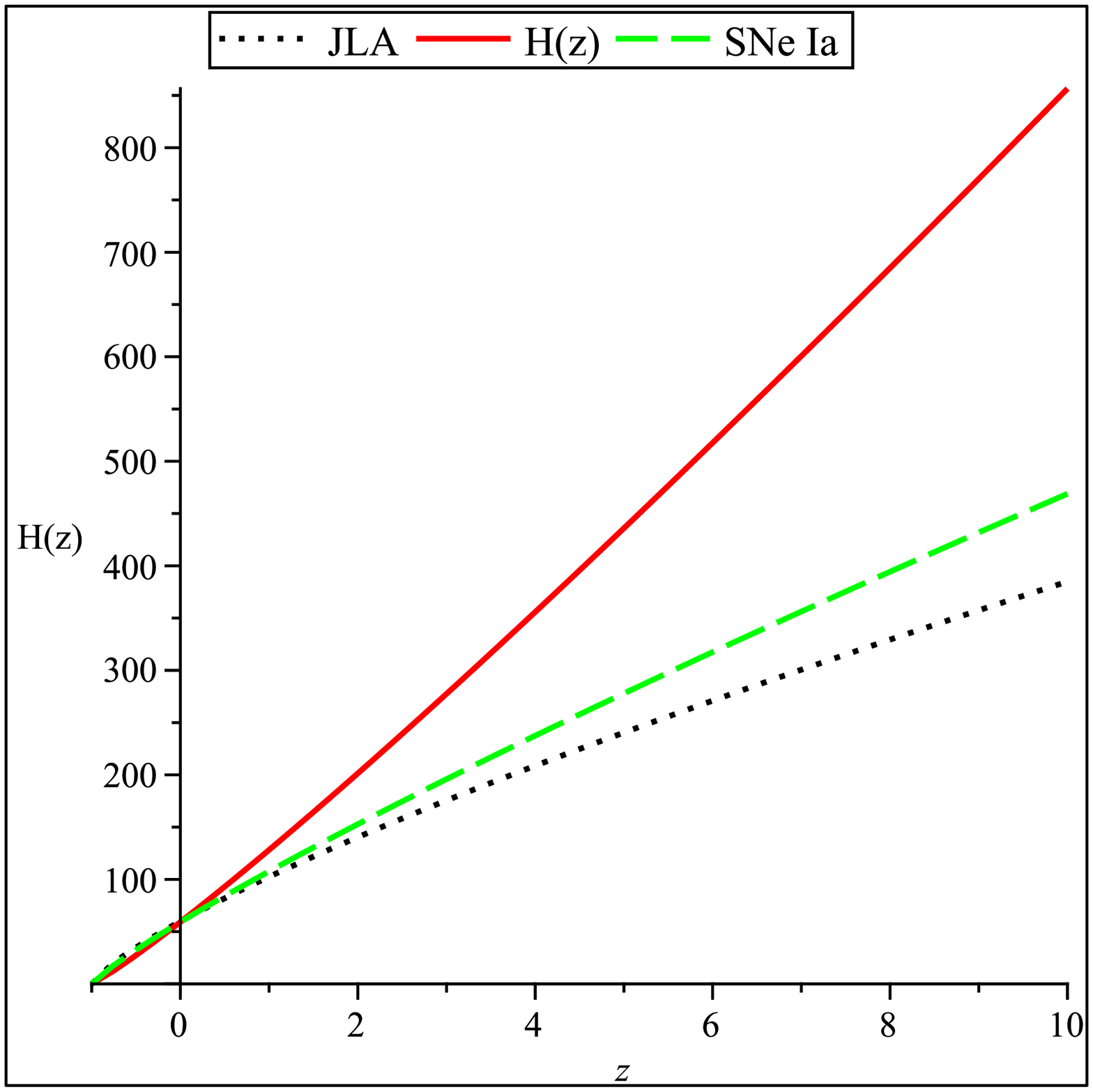}
	\caption{The plot of Hubble function $H(z)$ versus redshift $z$ for the best fit values of model parameters 
	$l$, $\lambda$, $\eta$ and $H_{0}$ in both cases of $\omega$.}
\end{figure}
%%%%%%%%%%%%%%%%%%%%%%%%%%%%%%%%%%%%%%%%%%%%%%%%%%%%%%%%%%%%%%%%%%%%%%%%%
Figures $1a$ \& $1b$ represent the best fit curve of the Hubble function $H(z)$ for the best fit values of the model parameters as 
well as the best fit values of the model parameters where the curvature parameter $k=-1,0,1 $ for the open, flat and closed models
displayed in Tables-$1$ \& $2$ for the three data sets as shown in the figures, respectively. Figures $2a$ \& $2b$ depict the behaviour 
of cosmological constant $\Lambda(z)$ over redshift $z$ and we see that it is an increasing function of $z$ which is consistent with 
observational findings. Figures $3a$ \& $3b$ show the age of the universe defined as $(t_{0}-t)=\int_{0}^{z}\frac{dz}{(1+z)H(z)}$ for 
both cases of $\omega$ and we have calculated the age of the present universe $t_{0}$ as $14.54821996\leq t_{0}\leq20.62684904$ Gyrs 
that is consistent with the observational results $t_{0}=13.6$ Gyrs.\\

%%%%%%%%%%%%%%%%%%%%%%%%%Figure 2 %%%%%%%%%%%%%%%%%%%%%%%%%%%%%%%%%%%%%%%%%%%%
\begin{figure}
	\centering
	a.\includegraphics[width=6cm,height=6cm,angle=0]{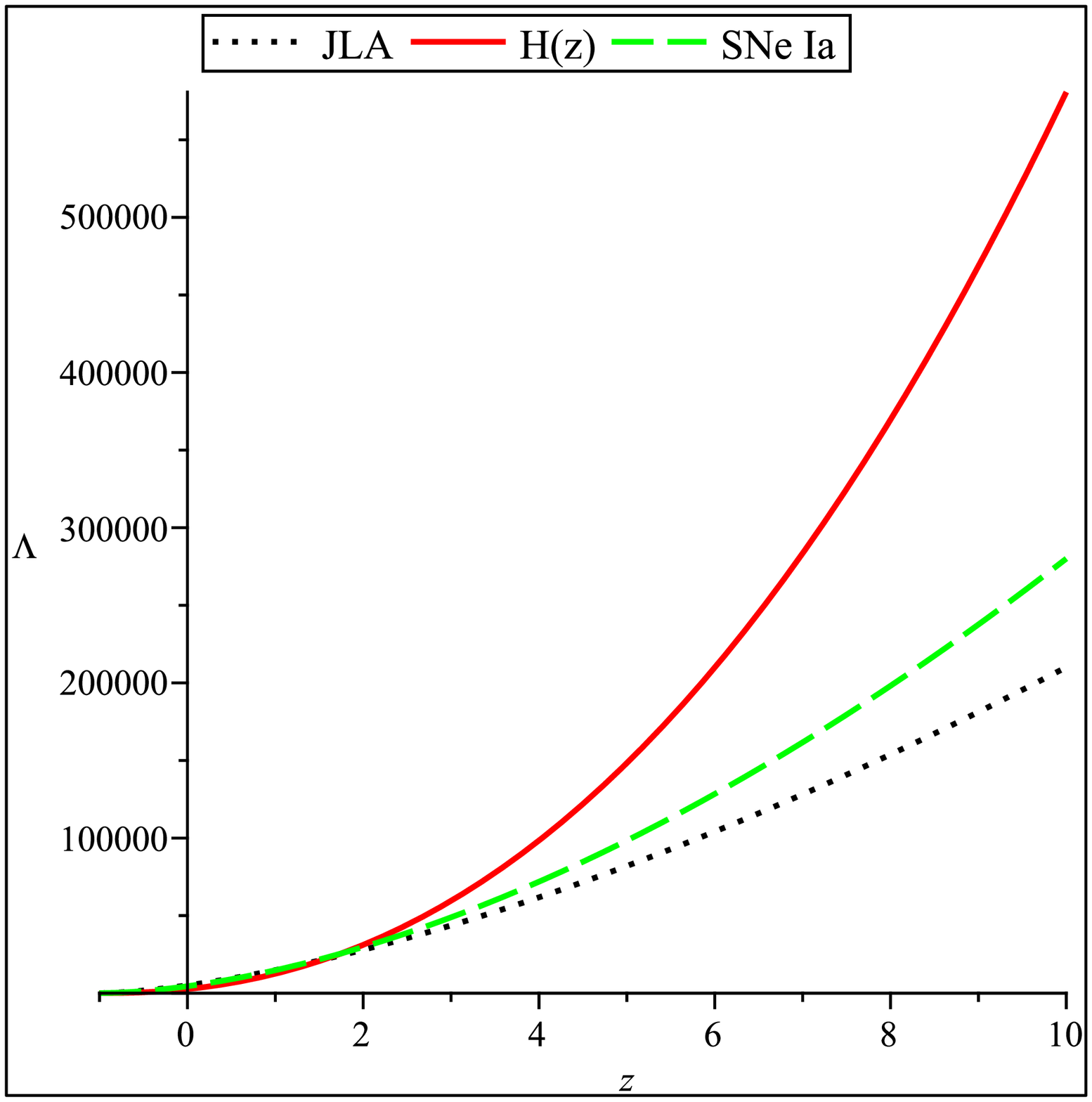}
	b.\includegraphics[width=6cm,height=6cm,angle=0]{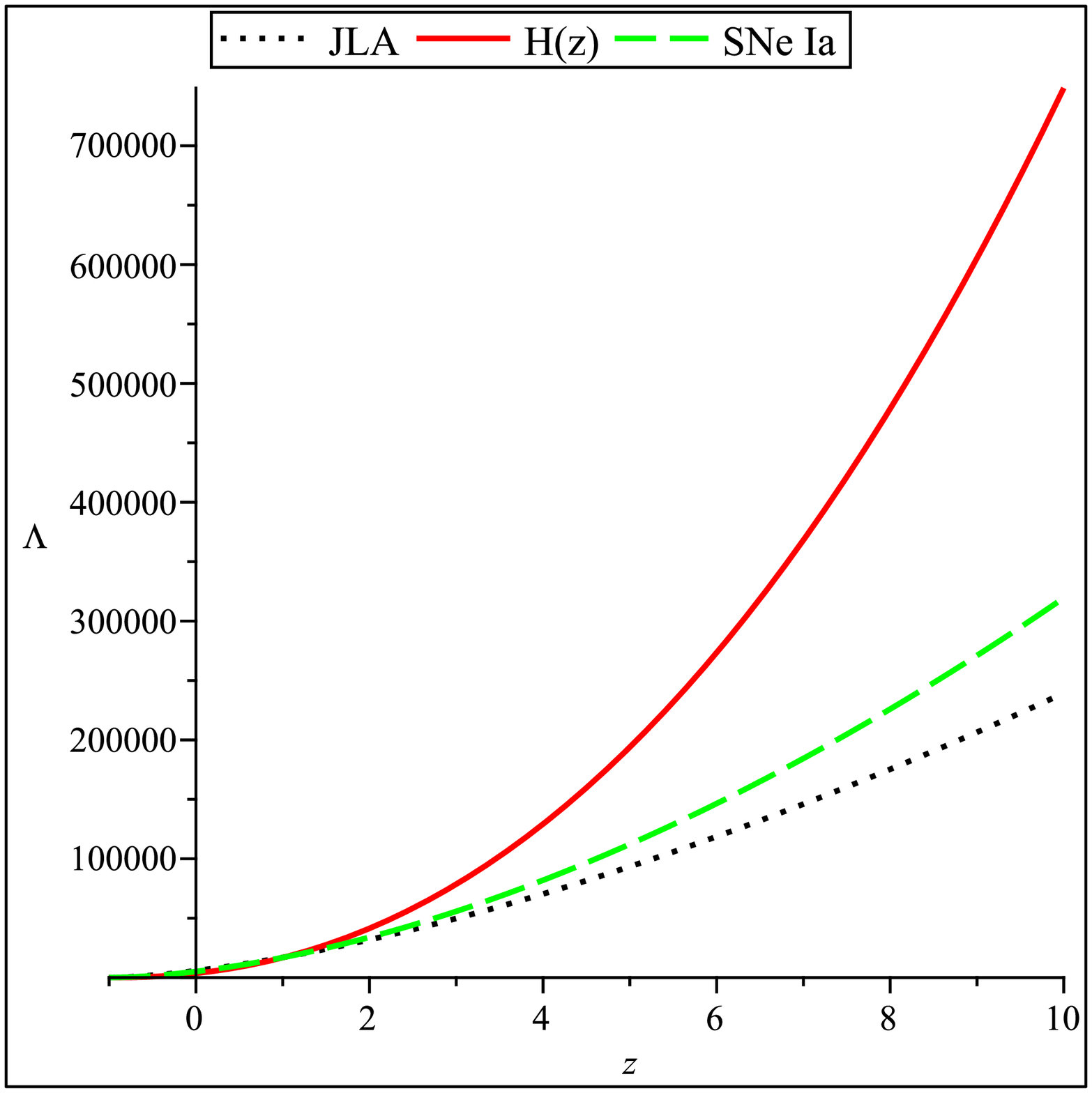}
	\caption{The plot of Cosmological constant $\Lambda(z)$ versus redshift $z$ for the best fit values of model parameters 
	$l$, $\lambda$, $\eta$ and $H_{0}$ in both cases of $\omega$.}
\end{figure}
%%%%%%%%%%%%%%%%%%%%%%%%%%%%%%%%%%%%%%%%%%%%%%%%%%%%%%%%%%%%%%%%%%%%%%%%%
%%%%%%%%%%%%%%%%%%%%%%%%%Figure 3 %%%%%%%%%%%%%%%%%%%%%%%%%%%%%%%%%%%%%%%%%%%%
\begin{figure}
	\centering
	a.\includegraphics[width=6cm,height=6cm,angle=0]{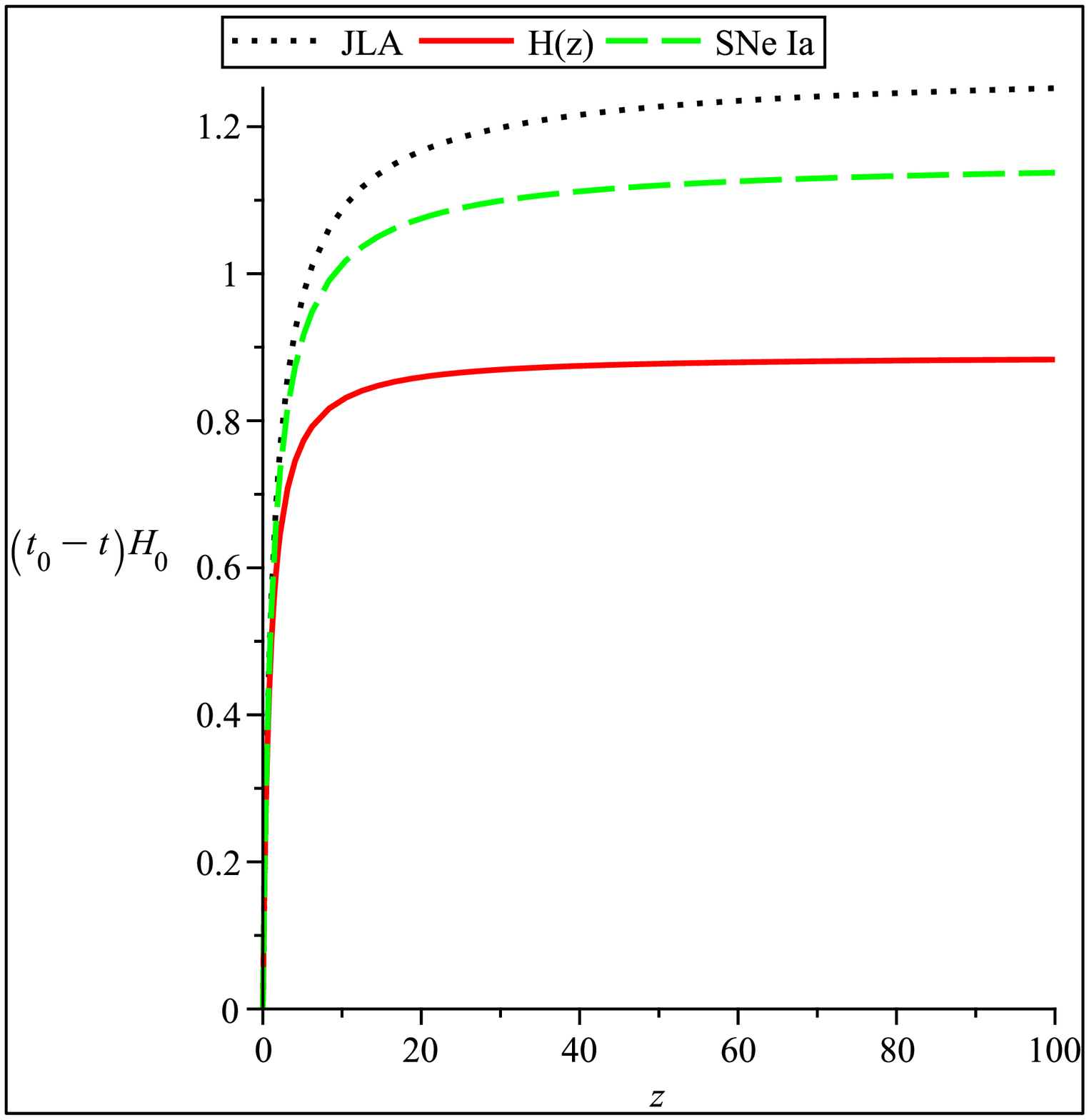}
	b.\includegraphics[width=6cm,height=6cm,angle=0]{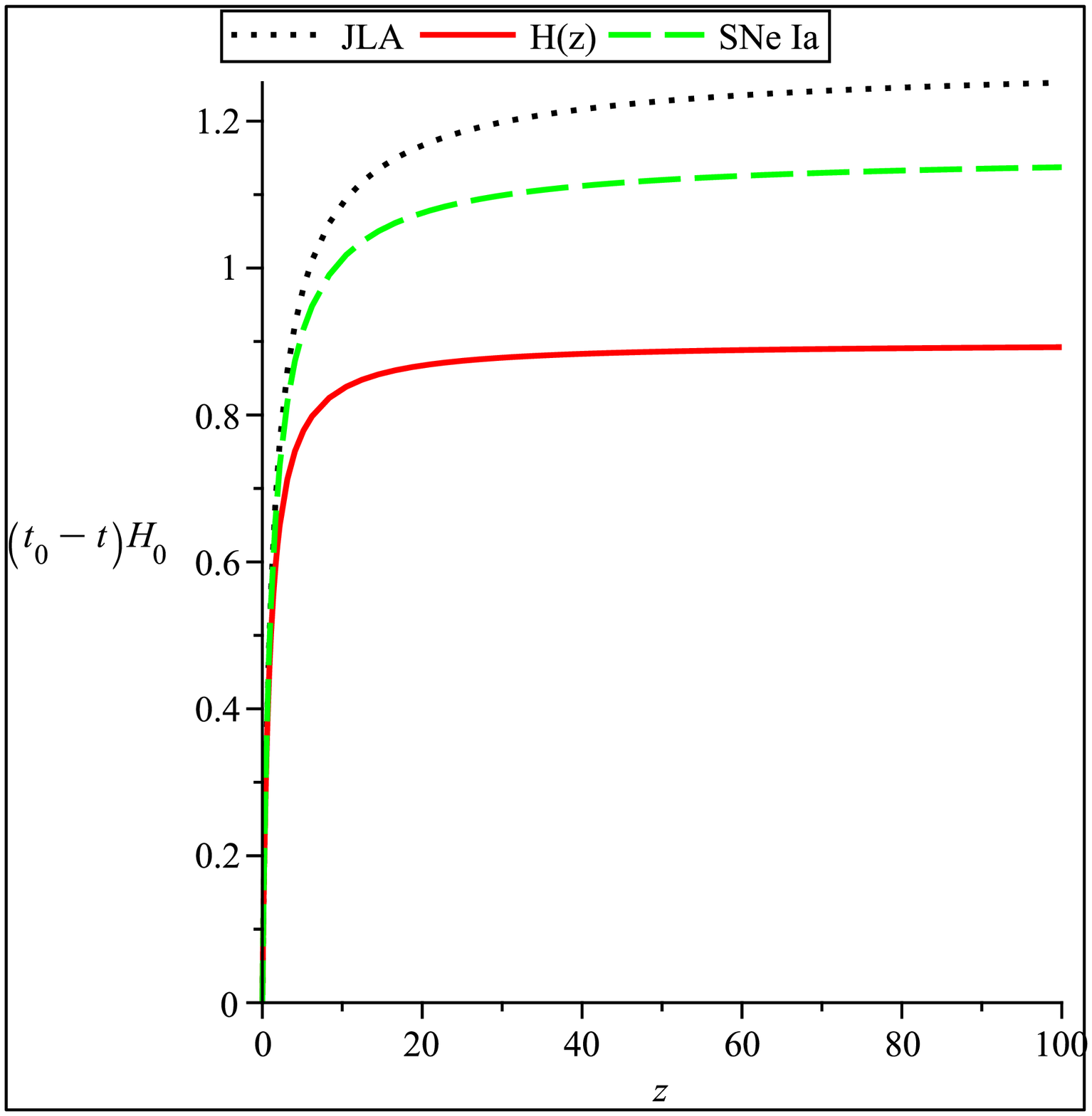}
	\caption{The plot of Cosmic time $(t_{0}-t)H_{0}$ versus redshift $z$ for the best fit values of model parameters 
	$l$, $\lambda$, $\eta$ and $H_{0}$ in both cases of $\omega$.}
\end{figure}
%%%%%%%%%%%%%%%%%%%%%%%%%%%%%%%%%%%%%%%%%%%%%%%%%%%%%%%%%%%%%%%%%%%%%%%%%

From Table-3 \& 4 we see that the value of total energy density parameter $0.9999999998\leq\Omega\leq1$. 
These findings are consistent with the observational constraints obtained by MAXIMA-1 flight and COBE-DMR experiments
\cite{ref32}, $\Omega=1.05\pm0.08 $from CBI-DMR observations \cite{ref18} and $\Omega=1.05\pm0.08 $on the total energy density 
parameter $\Omega=1.00^{+0.25} {-0.30}$. $\Omega=1.01\pm 0.03 $calculated from the first acoustic peak
in the CMB fluctuation angular power spectrum \cite{ref33}. \\

Moreover, we can obtain ranges of $l_0,\lambda_0$ and $\eta_0$ (the values of $l,\lambda, \eta$ in the present epoch) in 
the dust case ($w=0$) from the above equations. The first flight of the MAXIMA combined with COBE-DMR resulted in $0.25<\Omega_{m_0}<0.50$ 
and $0.45<\Omega_{\Lambda_0}<0.75$. Observations of SNeIa combined with the total energy density constraints from CMB and combined 
gravitational lens and steller dynamical analysis lead to $\Omega_{m_0}\sim0.3$ and $\Omega_{\Lambda_0}\sim0.7$. Sievers et al. 
and Sperger et al. obtained the values of $[\Omega_{m_0},\Omega_{\Lambda_0}]=[0.34\pm 0.12,0.67^{+0.10}_{-0.13}]$ and 
$[0.249^{+0.024}_{-0.031},0.719^{+0.021}_{-0.029}]$ respectively \cite{ref18,ref20}. Considering the varied observational dataset, 
many researchers prefer a range of $0.295\leq\Omega_{m_0}\leq0.365$ \cite{ref5}. \\

For dust case, from (\ref{omegam}) we obtain $3\Omega_m=2\frac{3-l-\lambda}{2+\eta-l}$. If we consider the case $\lambda=0=\eta$, 
we obtain $3.7937\leq l_0\leq 4.2099$. We note that using our model we are able to obtain the range of $l$ smaller than that of 
Arbab's \cite{ref34} which was $(3,4.5)$ and Ray's \cite{ref5} which was $[3.417,4.674]$. On a similar fashion we can find ranges of 
$1.905\leq \lambda_0\leq 2.115$ and $3.479\leq \eta_0\leq4.780$. So we see that using our model, we can generalize the previous results and 
also obtain the respective range of the parameters.

%%%%%%%%%%%%%%%%%%%%%%%%%%%%%%%%%%%%%%%%%%%%%%%%%%%%% SECTION 4 %%%%%%%%%%%%%%%%%%%%%%%%%%%%%%%%%%%%%%%%%%%%%%%%%%%%%%%%%%%%

\section{\textbf{Finite time singularities in the current model}}
We investigate equation (\ref{scalefac}) for possible finite time past or future cosmic singularities present in the current model. 
We consider distinct cases depending on the sign of $(2+\eta)-l(1+w)$.\\

\underline{Case 1}: $\frac{2+\eta}{l}>1+w$. 

If $(1+w)(3-l-\lambda)>0$, then there is a finite past singularity at $t_p=t_0-\frac{2+\eta-l(1+w)}{H_0((1+w)(3-l-\lambda))}<t_0$ 
whereas if $(1+w)(3-l-\lambda)<0$, then the scale factor diverges at some finite future time $t_f=t_0-\frac{2+\eta-l(1+w)}
{H_0((1+w)(3-l-\lambda))}>t_0$.\\

\underline{Case 2}: $\frac{2+\eta}{l}<1+w$. 

If $(1+w)(3-l-\lambda)<0$, then there is a finite past singularity at $t_p=t_0-\frac{2+\eta-l(1+w)}{H_0((1+w)(3-l-\lambda))}<t_0$ 
whereas if $(1+w)(3-l-\lambda)>0$, then the scale factor diverges at some finite future time $t_f=t_0-\frac{2+\eta-l(1+w)}{H_0((1+w)
(3-l-\lambda))}>t_0$. 
From this discussion it is clear that we can control the free parameters $l, \lambda, \eta$ to avoid any future singularity in the system. 

%%%%%%%%%%%%%%%%%%%%%%%%%%%%%%%%%%%%%%%%%%%%%%%%%% SECTION 5 %%%%%%%%%%%%%%%%%%%%%%%%%%%%%%%%%%%%%%%%%%%%%%%%%%%%%%%%%%%%%%%%%%%%%%%%%

\section{\textbf{Some cosmological parameters}}

The deceleration parameter $q=-\frac{a\ddot{a}}{\dot{a}^2}=-\frac{\dot{H}}{H^2}-1$ in cosmology is a dimensionless measure of the 
cosmic acceleration of the expansion of space. The statefinder parameters $(r,s)$ is defined originally by
$$r=\frac{\dddot{a}}{aH^3},\qquad s=\frac{r-1}{3(q-1/2)}.$$ 

We may link the $(r, s)$ parameters to the Hubble parameters $H$ and the deceleration parameter $q$ for our present model as 

\be r=2\left(\frac{\dot{H}}{H^2}\right)^2+3\frac{\dot{H}}{H^2}+1=q(2q+1)\label{r}\ee 

\be s=-\frac{2\dot{H}}{3H^2}=\frac{2}{3}(q+1)\label{s}.\ee

We also obtain a relation between $r$ and $s$ as 

\be r=\frac{9}{2}s(s-1)+1.\label{rs}\ee

In general, the statefinder hierarchy $A_n=\frac{a^{(n)}}{aH^n}$ for our present model can be calculated as
\bea A_{2k+1}=q(2q+1)(3q+2)\dots (2kq+2k-1),\notag\\
A_{2k}=-q(2q+1)(3q+2)\dots ((2k-1)q+2k-2).\eea

In this regard, it is to be mentioned that these handy relations between the various cosmological parameters as obtained in this 
section are valid for any time independent $w$ and with any of the decay laws considered in \cite{ref1,ref2,ref3,ref4} or 
any of their linear combinations.\\

For a $\Lambda$CDM model the statefinder pair ($r, s$) have the value ($1, 0$). In our present model we obtain this if and only if 
$\frac{\dot{H}}{H^2}=0$ which means the scale factor $a(t)$ increases exponentially with time. It also implies that $q=-1$, which is 
true if $\alpha+\beta=3$, but this condition also provide $\Lambda$ is truly a constant. \\

\textbf{Dust Case:}\\

Below we obtain some important results for the particular case $w=0$.\\

First and foremost, we have\be q=\frac{3-l-\lambda}{2+\eta-l}-1.\ee 
Hence for an accelerating universe ($q<0$) we obtain a relation $\lambda+\eta>1$. \\

We notice that for dust case, the standard model formula $\Omega_m = 2q$ is now replaced by $\Omega_m = \frac{2}{3}q + \frac{2}{3}$, 
same as the result of Arbab \cite{ref4}. However, both models give $q = 1/2$
for a critical density $\rho_c=\frac{3H^2}{8\pi G}$. We also obtain $s=\Omega_m$.\\
 
Furthermore, using equations (\ref{hdot}) and (\ref{omegalambda}) we obtain that the deceleration parameter $q$ and the vacuum density 
parameter $\Omega_\Lambda$ are connected by the relation 

\be3\Omega_\Lambda=1-2q+2w\frac{3-l-\lambda}{(2+\eta)-l(1+w)},\ee

and so for dust case, an accelerating universe $(q<0)$ requires $\Omega_\Lambda>1/3$, which perfectly fits the modern observational 
data as the present accepted value of $\Omega_\Lambda$ is 0.7, much larger than 1/3. Hence our model fits an accelerating universe.

%%%%%%%%%%%%%%%%%%%%%%%%%%%%%%%%%%%%%%%%%%%% SECTION 6 %%%%%%%%%%%%%%%%%%%%%%%%%%%%%%%%%%%%%%%%%%%%%%%%%%%%%%%%%%%%%%%%%

\section{\textbf{Conclusion}}

We have considered a spatially homogeneous and isotropic spacetime in the presence of a dynamic cosmological term $\Lambda(t)$ satisfying 
$\Lambda=l(\frac{\dot{a}}{a})^2+\lambda\frac{\ddot{a}}{a}+4\pi G\eta\rho$ motivated by modification of the Einstein-Hilbert action. 
This generalizes several results of the previous models in flat FRW spacetime and also enables us to find the ranges for $l, \lambda, \eta$ 
from the observational data apart from the interrelation we obtain between these parameters for accelerating universe ($q<0$). We explore 
the recently defined statefinder hierarchy for this generalized model of ours and prove some effective results for any time independent $w$. 
Furthermore, we show that our model fits in perfectly with the modern observational datasets.\\

We have found that the scale factor $a(t)$ vanishes at $t=t_0-\frac{\{(2+\eta)-l(1+w)\}}{H_0(1+w)(3-l-\lambda)}$ while for this 
value of time the Hubble function $H$ and the cosmological term $\Lambda$ get infinitely large values. We have found that with increasing 
time or decreasing redshift the scale factor increases but the $H$ and $\Lambda$ decreases to a finite small values in late time universe. 
These behaviours of cosmological parameters reveal that our universe model starts with a finite time big-bang singularity stage and goes 
on expanding till late time. We have also noticed that the total energy density parameter $\Omega\leq1$ which indicates towards the closed 
and flat geometry of the universe and it was compatible with several observational results. We have also found a constraint on dark energy 
density parameter $\Omega_{\Lambda_0}>\frac{1}{4}$ to provide an acceleration in expansion of the universe which is compatible with the 
result $\Omega_{\Lambda_0}>\frac{1}{3}$. We have also calculated the age of the present universe in the range $14<t_{0}<21$ Gyrs.
%%%%%%%%%%%%%%%%%%%%%%%%%%%%%%%%%%%%%%%%%%%%%%%%%%%%%%%%%%%%%%%%%%%%%%%%%%%%%%%%%%%%%%%%%%%%%%%%%%%%%%%%%%%%%%%%%%%%%%%
\section*{Acknowledgments} 
 De Avik and Tee How Loo are supported by the grant FRGS/1/2019/STG06/UM/02/6. A. Pradhan would like to express his
appreciation to the Inter-University Centre for Astronomy \& Astrophysics (IUCAA), India for supporting under the visiting associateship.

%%%%%%%%%%%%%%%%%%%%%%%%%%%%%%%%%%%%%%%%%%%%%%%%%%%%%%%%%%%%%%%%%%%%%%%%%%%%%%%%%%%%%%%%%%%%%%%%%%%%%%5

\end{document}